\documentclass[pra,twocolumn,amsmath,amssymb,superscriptaddress,unsortedaddress,floatfix]{revtex4}
\usepackage{graphicx}% Include figure files
\usepackage{dcolumn}% Align table columns on decimal point
\usepackage{bm}% bold math
\usepackage{color}
\usepackage{bbm}
\usepackage{float}
\usepackage{bbold}

\usepackage{epstopdf}

\usepackage{soul}

\def\dt{{\rm d \!}}

\def\<{\langle}
\def\>{\rangle}
\def\be{\begin{equation}}
\def\ee{\end{equation}}
\def\tr{\rm Tr}
\def\bei{\begin{itemize}}
\def\eei{\end{itemize}}

\usepackage[a4paper, left=2cm, right=2cm, top=2cm, bottom=2cm, headsep=1.2cm]{geometry}

\begin{document}
\title{Sharp transitions in low-number quantum dots Bayesian magnetometry}

\author{Pawe{\l}  Mazurek*}

\affiliation{Institute of Theoretical Physics and Astrophysics, National Quantum Information Centre, Faculty of Mathematics, Physics and Informatics, University of Gda\'nsk, Wita Stwosza 57, 80-308 Gda\'nsk, Poland}

\email{pawel.mazurek@ug.edu.pl}

\author{Micha{\l} Horodecki}

\affiliation{Institute of Theoretical Physics and Astrophysics, National Quantum Information Centre, Faculty of Mathematics, Physics and Informatics, University of Gda\'nsk, Wita Stwosza 57, 80-308 Gda\'nsk, Poland}

\author{\L{}ukasz Czekaj}
\affiliation{Institute of Theoretical Physics and Astrophysics, National Quantum Information Centre, Faculty of Mathematics, Physics and Informatics, University of Gda\'nsk, Wita Stwosza 57, 80-308 Gda\'nsk, Poland}

\author{Pawe{\l} Horodecki}

\affiliation{Faculty of Applied Physics and Mathematics, Gda\'nsk University of Technology, Narutowicza 11/12, 80-233 Gda\'nsk, Poland}
\affiliation{National Quantum Information Centre in Gda\'nsk, Andersa 27, 81-824 Sopot, Poland}

\begin{abstract}
We consider Bayesian estimate of static magnetic field, characterized by a prior Gaussian probability distribution, in systems of a few electron quantum dot spins interacting with infinite temperature spin environment via hyperfine interaction. Sudden transitions among optimal states and measurements are observed. Usefulness of measuring occupation levels is shown for all times of the evolution, together with the role of entanglement in the optimal scenario. For low values of magnetic field, memory effects stemming from the interaction with environment provide limited metrological advantage.
\end{abstract}

\maketitle	

\section{INTRODUCTION}
Quantum metrology relies on the fact that quantum correlations make state evolution more sensitive to dynamics which depends on some parameter that is supposed to be revealed. It is known that, in the so called frequentist approach, for estimating small variations of a deterministic parameter, for locally unbiased estimators dependent on its value and $N$ systems undergoing independent evolution, quantum mechanics can offer a $1/N$ (so called Heisenberg scaling) improvement of the precision (defined by the deviation from the precise value) in the asymptotic limit \cite{Braunstein1992, Braunstein1994, Giovannetti2004}. This should be compared to a scaling $1/\sqrt{N}$, available for classical resources, and referred to as quantum shot-noise limit. Generally it is known that in a situation when the parameter is a phase generated by some Hamiltonian evolution, then the local noise usually destroys the quantum effect (both in atomic spectroscopy \cite{Huelga1997, Shaji2007} and quantum optics \cite{Kolodynski2010, Knysh2011}), leading to at most constant improvement over classical scaling. 

In the so-called Bayesian approach \cite{Helstrom1976, Trees2007, Macieszczak2014}, this scenario is altered so that the parameter to be estimated is a random variable with some a priori probability distribution. In many cases, this framework is more justified than the frequentist approach: it does not assume perfect knowledge about a system under consideration before an experiment and it outputs optimal estimators even for small N. Compared to quantum frequentist approach, there is not much work regarding quantum Bayesian approach. In particular, precision bounds in the asymptotic limit have been established \cite{Tsang2011,Tsang2012}. In \cite{Macieszczak2014}, authors investigate effects of noise introduced as a classical random phase, and \cite{Demkowicz2010} takes into account photonic losses. However, in both cases the noise does not depend on the value of the parameter to be estimated.
 
In this paper we apply Bayesian metrology to a physical scenario where the form of the noise depends on the parameter. Specifically, we analyze a system of independent quantum dots interacting via hyperfine interaction with their local, maximally mixed spin environments \cite{Mazurek2014b}, under a so called box model approximation \cite{Barnes2011,Merkulov2002}. Spins of the electron dots are subject to external time independent magnetic field $B$ with the random value according to Gaussian probability distribution with a given variance  $\Delta^2 B_{prior}$ and mean $B_{0}$. As a case study, we choose a Gaussian prior due to its unimodality and the fact that it does not vanish for all possible values of the parameter to be estimated, which describes well possible initial knowledge about the unbounded parameter. The Bayesian approach allows to diminish the variance of magnetic field estimator which relies on measurements that may in general depend on time and give extra information about that value. The purpose of our analysis is twofold: (i) We want to see to what extent entangled states allow for better estimation than classical resources.
(ii) We want to check how the sensitivity of the dynamics depends 
on initial states for different times of duration of the evolution. 
Regarding the first goal, we are looking for such time duration, for which there is an optimal reduction of relative uncertainty. In the second case, 
we look for an optimal reduction of uncertainty at each time, and analyse
the form of states leading to such optimal reduction. 

We find that, in the considered model, entanglement wins over classical estimation, however only to a very modest extent. More importantly, the analysis of an optimal estimation at given times reveals sharp transitions in the space of optimal initial states and optimal observables. In particular, we can identify transitions of the zeroth kind (when there is a discontinuity in the spectrum of an optimal observable), and of the first kind (with discontinuity in the first derivative of the spectrum of an optimal observable). Due to rapidly increasing numerical complexity, our analysis is restricted to low number of dots. However, the mentioned sharp transitions appear irrespectively of the value of N, with the number of transitions increasing with N.

\section{OUTLINE OF THE RESULTS}\label{sec:outline}

In this section we give an overview of the main results. The more detailed description will be given in Sec. Optimal Strategies for Magnetometry. Our aim is to find the optimal initial state and measurement scheme which results in the narrower variance $\Delta^{2} B_{est}$ -- a signature of the gain of information about the field. It is achieved by numerical optimization \cite{Macieszczak2014} yielding optimal strategies for given time of the evolution and initial probability distribution of the parameter. 

For a given variance and mean, after initial time of approximately unitary evolution, a transition of the first kind occurs –- the optimal observable changes such that its spectrum attains a non-differentiability point during the transition. The optimal initial state changes smoothly from GHZ state of  N  qubits, $\frac{1}{\sqrt{2}}\big(|0\>^{\otimes N}+|1\>^{\otimes N}\big)$, to a coherent superposition of GHZ with a product of 'plus' states, $| +\rangle^{\otimes N}$. The GHZ-plus superposition is the one that achieves the global optimum in the rate of $\frac{\Delta^2 B_{est}}{\Delta^{2} B_{prior}}$.  The evolution of the optimal strategy is further marked by transitions of the zeroth kind, they are accompanied by non-smooth transitions in the space of initial states. The optimal initial state jumps to $|+\rangle\otimes\dots\otimes|+\rangle\otimes|0\rangle$, then to $|+\rangle\otimes\dots \otimes|+\rangle\otimes |0\rangle\otimes |0\rangle$, 
and so on. Thus, the number of qubits that should be prepared in $|+\rangle$ states decreases with increasing time of measurement, until the product structure with no coherences in the standard basis becomes optimal for long times. 

The above can be explained by the interplay between two metrological strategies, out of which the first relies on information extracted from the field-dependent phase factors, and the second is based on field-dependent population levels. Therefore, optimizing over initial states and measurement strategies can be rephrased in terms of searching for a balance between these two strategies. The large number of the observed transitions is the consequence of this trade-off evaluated for different times of the evolution. In the most global picture, we find quantum strategies relying on initial entanglement be superior to classical ones in the first stages of the evolution (where the global optimum is achieved), while product, non-coherent strategies are dominant in long time regime. This is in this regime that, for a unitary evolution, solely phase-dependent strategies would lead to no increase of information about the magnetic field value, due to a non-zero $\Delta^{2} B_{prior}$.

However, one should note that measurement strategies in all the regimes rely at least partially on measurements of occupation levels, in a response to the noisy character of the quantum channel. For small times, GHZ remains the optimal state, and optimal measurement scheme includes projections onto occupation levels initially non-available by the state, which are output with probabilities raising with time. Another metrological feature of the system is that the population scenario is {\it not} the only one that could be used for long times. As a signature of non-Markovian evolution, in small fields we observe a partial recovery of the initial correlation factors of spin states, with information about the magnetic field encoded in their amplitude.

\section{MODEL OF AN ELECTRON SPIN IN A QUANTUM DOT SYSTEM} \label{sec:dots}
Below we describe the system of an electron confined within a quantum dot,  interacting with an environment of nuclear spins. In the so called box-model \cite{Mazurek2014b}, the  Hamiltonian (in units $\hbar$=1) of a dot is given by 
\begin{equation}
\Omega S^{z} + \alpha S^{z} I^{z}+\frac{\alpha}{2}(S^{+}I^{-}+S^{-}I^{+}),\end{equation} 
with $\Omega=-g\mu_{B}B$, where $B$ denotes the magnetic field, $\mu_{B}$ is a Bohr magneton, and $g$ an effective gyro-magnetic factor of an electron in a dot. Above we use the operator of the total nuclear spin $I=\sum_{k}I_{k}$ and its projection $I^{z}$ onto the direction of the magnetic field, with eigenstates $|K,m\rangle$ and associated eigenvalues $\sqrt{K(K+1)}$ and $m$, respectively. In the box model, we assume that hyperfine interaction constant is the same for all nuclei inside a quantum dot and equal to $\alpha=\mathcal{A}/n$, where $n$ is the number of nuclei interacting with the electron spin, and $\mathcal{A}$ is the value of total hyperfine interaction energy dependent on a host material. Our calculations are performed for quantum dots in gallium arsenide, where we have $n= 1.5 \times 10^{6} $ and $\mathcal{A}=83$ $\mu$eV \cite{Liu2007, Barnes2011, Mazurek2014b}. Furthermore, with a requirement that $\mathcal{A}/n$ remains constant, we simulate spin evolution with small numbers of nuclei in the environment. This is due to disappearance of few-body coherent effects already for environments composed of 10 nuclei, and the fact that 50 spin systems are large enough to reproduce large-number-of-nuclei evolution \cite{Mazurek2014b}. General conclusions of the paper hold for quantum dot systems in every material with hyperfine interaction playing a leading role in decoherence of electron spin, as long as decoherence takes place within the box model time range of validity. The box model of hyperfine interaction is valid for initial times of the evolution wherever the state of environment is maximally mixed. The time range of applicability of the box model is $t\ll n/\mathcal{A}$ \cite{Barnes2011}, which is $1.2\times 10^{4}$ ns for the parameters used. Unless the system is especially experimentally prepared, for quantum dots in gallium arsenide the state of environment can be taken as maximally mixed due to small values of nuclear gyro-magnetic ratios, which results in nuclear Zeeman splittings corresponding to less than a mK for each Tesla of magnetic field applied to the system.   

In the basis of total nuclear spin and its projection to the direction of the magnetic field, the state of environment can be therefore expressed as  $\rho_{env}=\sum_{K,m} P_{K,m}|K,m\rangle\langle K,m|$ with $\{P_{K,m}\}$ satisfying $\sum_{K,m}P_{K,m}=1$ and 
\begin{equation}
P_{K,m}\sim \sum_{i}(-1)^{i}{n\choose i}{(s+1)n - (2s+1)i-K-2 \choose n-2},
\end{equation} 
where $i\in[0,n]$ is an integer \cite{Mihailov1977}.

We obtain the following form of a single-dot evolution
\be\label{reduced_ev0olution}
\rho(t)=\sum_{i=1}^{4}K_{i}\rho(0)K^{\dagger}_{i}, 
\ee
with Kraus operators, satysfying $\sum_{i} K^{\dagger}_{i}K_{i}=\mathbb{1}$ and $K_{i}>0$, given by
\begin{equation}\label{K1}
K_{1}=\sqrt{1-A}|0\rangle\langle 1|,
\end{equation}
\begin{equation}\label{K2}
K_{2}=\sqrt{1-A}|1\rangle\langle 0|,
\end{equation}
\begin{equation}\label{K3}
K_{3}=\frac{1}{\sqrt{2}}\sqrt{A+|E|}\Big[  \frac{|E|}{\overline{E}}|0\rangle\langle 0|+|1\rangle\langle 1|  \Big].
\end{equation}
\begin{equation}\label{K4}
K_{4}=\frac{1}{\sqrt{2}}\sqrt{A-|E|}\Big[  -\frac{|E|}{\overline{E}}|0\rangle\langle 0|+|1\rangle\langle 1|  \Big].
\end{equation}

We use the notation $S^{z}|0\rangle=\frac{1}{2}|0\rangle$, $S^{z}|1\rangle=-\frac{1}{2}|1\rangle$, $S^{\pm}=S^{x}\pm i S^{y}$ for electron spin operators $S$ and similarly for total spin operator $I$ of the nuclei. For brevity, above we do not show dependence on $B$, $t$ and $\alpha$ of the following terms:  
\begin{equation}
A(B,t,\alpha)=\sum_{Km}P_{K,m}|X_{K,m}(t)|^2,
\end{equation}

\begin{equation}
E(B,t,\alpha)=\sum_{Km}P_{K,m}X_{K,m}(t)X_{K,m-1}(t),
\end{equation}

\begin{equation}
X_{K,m}(B,t,\alpha)=\cos^{2}\theta_{K,m} e^{-i\chi_{K,m} t}+\sin^{2}\theta_{K,m} e^{+i\chi_{K,m} t},
\end{equation}

\begin{equation}
\chi_{K,m}(B,\alpha)=\frac{1}{2}\sqrt{\Omega^{2}+\Omega\alpha(2m+1)+\frac{\alpha^{2}}{4}+\alpha^{2}K(K+1)}.
\end{equation}

We have 
\begin{equation}
\sin{\theta_{K,m}}=\frac{M_{K,m}}{(E^{+}_{K,m}+E_{m+1})^{2}+M^{2}_{K,m}}, 
\end{equation}
with $E_{m}(B,\alpha)=\frac{\Omega+\alpha m}{2}$, $M_{K,m}(B,\alpha)=\frac{\alpha}{2}\sqrt{K(K+1)-m(m+1)}$, $E^{+}_{K,m}=-\frac{\alpha}{4}+\frac{\sqrt{(E_{m}+E_{m+1})^{2}+4M^{2}_{K,m}}}{2}$. 

For very large magnetic fields ($B>\frac{n\alpha}{g\mu_{B}}$), distribution in the environment of the Zeeman energy $\Omega$ (resulting from a spin flip)  is suppressed. This leads to freezing of occupation levels ($A(t)=1$), and pure dephasing of coherences ($E(t)=\exp(i g\mu_{B}Bt)\exp(-\big(\frac{t}{T_{2}^{*}}\big)^{2})$, with $T_{2}^{*}=\sqrt{\frac{6}{I(I+1)n}}/\alpha$). For smaller magnetic fields, oscillations of occupation levels appear. For very small magnetic fields ($B<\frac{\sqrt{n}\alpha}{g\mu_{B}}$), phase decoherence resembles the evolution of occupation levels, which are partially leveled out in high evolution times due to the interaction with environment.

\section{BAYESIAN METROLOGY}\label{sec:metrology}

In this section we will recall Bayesian approach from \cite{Macieszczak2014}. We consider a parametrized family of states $\{\rho_B\}$, and assume some prior distribution $p(B)$ over the parameter $B$, which in our case will be the magnetic field. 
Denote by $B_0$ the average value of  $B$ according to prior distribution $B_0=\int \dt B  B p(B)$.
In order to estimate the value of $B$, one performs a measurement $\{\Pi_{\tilde B}\}$, 
whose outcomes $\tilde B$ are the estimated values of $B$. After the measurement, knowledge of the parameter $B$
is supposed to increase. 
The resulting variance (averaged over the distribution of the magnetic field) is defined by 
\be \label{eq:new_variance}
\Delta^2 B _{est} = \int \dt B \dt \tilde B p(B,\tilde B) (B - \tilde B)^2,
\ee
where a joint probability distribution $p(B, \tilde B)$ is given by 
\be
p(B,\tilde B)= p(B) \tr(\rho_B \Pi_{\tilde B}). 
\ee

One is interested in a measurement which maximizes the difference between the variance of the prior $\Delta^2 B_{prior}$
and a post-measurement variance $\Delta^2 B_{est}$. It is proven \cite{Personick1971} (see an easy proof in \cite{Macieszczak2014}) 
that the optimal measurement is a von Neumann one, so that POVM elements $\Pi_{\tilde{B}}$ are orthogonal projections,
and the measurement can be represented by an observable $L=\sum_{\tilde B} \tilde B \Pi_{\tilde B}$.
Then the maximal difference is given by 
\be
\Delta^2 B_{prior} - \Delta^2B_{est} = {\tr} (\bar \rho L^2) - B_0^2,
\ee
where $\bar \rho = \int \dt B p(B) \rho_B$ and the optimal observable $L$ can be computed from 
\be
\frac12 (L \bar \rho + \bar \rho L)= \bar \rho',
\ee
with 
\be
\bar \rho'= \int \dt B p(B) \rho_B B. 
\ee
Explicitly, $L$ is given by 
\be
L=2\sum_{i,j} \frac{|i\>\<i|\bar \rho' |j\>\<j|}{\lambda_i + \lambda_j}
\label{eq:L}
\ee
where $\bar \rho=\sum_{i}\lambda_i|i\>\<i|$.

We shall exploit the above in the next section.
For the sake of slightly broader view let us recall
briefly a connection of the present picture with
Fisher information, the quantity that belongs to the asymptotic regime.
Namely, under some regularity conditions \cite{Gill1995}, the following bound  holds
:
\be
\Delta^2B_{est}\geq \frac{1}{\int \dt B p(B) F_{B} +\mathcal{I}}, 
\ee
where $F_{B}=\tr (\rho_{B} L^{2}_{B})$ is quantum Fisher information, determined   
the symmetric logarithmic derivative operator $L_{B}$ defined by  
\be
\frac12 (L_{B} \rho_{B} + \rho_{B} L_{B})= \rho'_{B}, 
\ee
where $\rho'_{B}=\frac{\rho_{B}}{\dt B}$ and $\mathcal{I}=\int \dt B \frac{1}{p(B)}\big(\frac{\dt p(B)}{\dt B}\big)^{2}$. For a Gaussian distribution one has $\mathcal{I}=\frac{1}{\Delta^2B_{prior}}$.

\section{OPTIMAL STRATEGIES FOR MAGNETOMETRY}\label{sec:results}
\begin{figure*}
\includegraphics[scale=0.73]{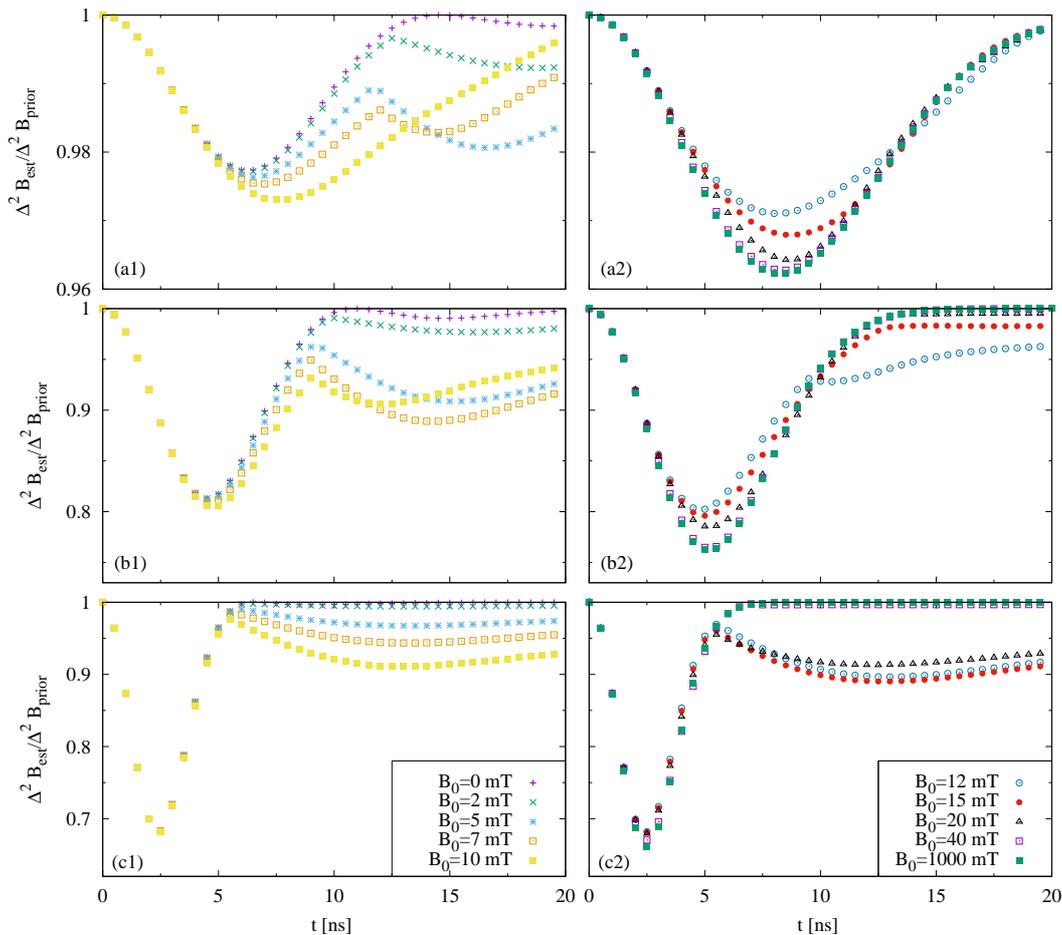}
\caption{\label{fig:1qubit_Delta} Estimation of $B$ for a single qubit. We plot $\Delta^2B_{est}/ \Delta^2B_{prior}$ optimized over states,
against time, for several values of the mean $B_0$ and variance $\Delta B_{prior}=1$ mT (a1, a2), $\Delta B_{prior}=4$ mT (b1, b2), $\Delta B_{prior}=10$ mT (c1, c2).   
For small $B_0$ there are two time regimes, the phase one to the left and the population one to the right.} 
\end{figure*}

In our model of the evolution of a system of $N$ independent quantum dots, the family of states $\{\rho_{B}\}$ at specified time $t$ is determined by the magnetic field $B$, time $t$ and an initial N-qubit state $\rho^{N}(0)$ by
\begin{eqnarray}
\rho_B(t)\nonumber=\Lambda_{B,t}(\rho^{N}(0))=\nonumber\\
\sum_{i_{1},\dots,i_{N}=1}^{4}K_{i_{1}}\otimes\dots\otimes K_{i_{N}}\rho^{N}(0)K_{i_{1}}^{\dagger}\otimes\dots\otimes K_{i_{N}}^{\dagger},
\end{eqnarray}
where $\{K_{i}\}_{i=1,\dots,4}$ is given by Eq. (\ref{K1}) - (\ref{K4}). We are interested in minimizing the 
$\Delta^2 B_{est}$ of Eq. (\ref{eq:new_variance}) over time and initial states, for a given prior distribution.
The latter we take Gaussian, determined by an average $B_0$ and a variance $\Delta^2 B_{prior}\equiv \Delta^2 B$. 
Our aim, among others, is to understand to what extent multi-particle entanglement can lead to a smaller value of $\Delta^2 B_{est}$. 

For optimizing over all initial states, we will use an iterative algorithm of \cite{Macieszczak2014}. At the first instance, one picks a random initial state. Then the optimal observable $L$ is computed according to the formula \eqref{eq:L}. An eigenvector of $\int \dt B p(B) \Lambda^{*}_{B}(L^{2}-2BL)$ corresponding to its smallest eigenvalue is taken for the next iteration. In the above, $\Lambda^{*}_{B}$ corresponds to a map dual to $\Lambda_{B}$. One continues the iterations until obtaining a convergence in $\Delta^{2}B_{est}$. Independent runs of the above algorithm enable the calculation of the optimal preparation and measurement strategy for given values of $N$, $B_{0}$, $\Delta^{2}B_{prior}$ and $t$.

{\it Single qubit (N=1)}We start with the analyzis for one quantum dot. In Fig. \ref{fig:1qubit_Delta} we depict the ratio $\Delta^2B_{est}/ \Delta^2B_{prior}$, optimized over all initial states. For any fixed spread $\Delta^2B_{prior}$, 
we consider several values of the mean $B_0$.  For large fields, we observe single minimum, while for small fields there are two minima. 
To understand this difference, let us first consider the large fields.
For the large fields, occupation levels are not influenced by the dynamics, hence an estimation of the magnetic field $B$ can only be done through the phase dependence on $B$.
The single minimum comes from a trade-off between damping of the phase, and rotation of the phase by the magnetic field. For times long enough so that the coherences are nearly completely damped, the state ceases to depend on the magnetic field, hence there is no improvement, i.e.  the variance after estimation is equal to the variance of the prior distribution. As in the noiseless case, a representation of the optimal state on a Bloch sphere is perpendicular to the direction of the field, and 
the representation of the optimal observable is perpendicular to both the field and the state directions. Indeed, in this way, the state is most sensitive to changes of the field, and also the observable is most sensitive to changes of the state during the evolution. 

\begin{figure}
\includegraphics[scale=0.62]{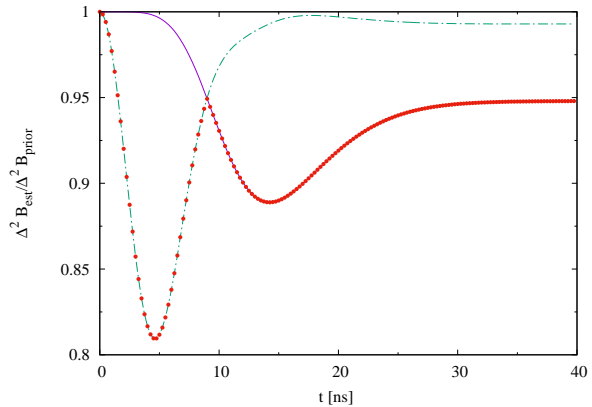}
\caption{\label{fig:1dot_regimes} Comparison between 'perpendicular' (green dashed line) and 'parallel' (solid purple line) strategies for 1 quantum dot and $B_{0} = 7$ mT, $\Delta B_{prior}=4$ mT. Red points represent the optimal strategy.} 
\end{figure}

Let us now consider the small fields. Here the situation becomes more complex as the dynamics affects both coherences as well as the diagonal of the density matrix of the electron, and the change of the diagonal depends on $B$. We see a cusp dividing two regions, each one possessing its minimum. The interpretation is the following: We have two time regimes, phase regime and population regime. In the phase regime, the phases are not yet damped, and the mechanism of the estimation is based as before on 
a rotation of the state on a Bloch sphere in a plane perpendicular to the direction of the field. In the population regime, the phases are damped,
and the estimation is based on measuring occupation levels. In this regime both the optimal state and the optimal observable are parallel to the field direction, as in this way, the populations will be most affected by field changes, and the parallel observable is just the one that measures occupation levels. The cusp between the two regimes is formed by an intersection of curves corresponding to 
the aforementioned "perpendicular" and "parallel" strategy,  with the optimal states $\frac{1}{\sqrt2}|0\> + |1\>$ and $|0\>$. This is visible in Fig. \ref{fig:1dot_regimes}, where the evolution of $\frac{\Delta^2 B_{est}}{\Delta^2 B_{prior}}$ is showed for $B_{0}=7$ mT and $\Delta B_{prior}=4$ mT. One should note that for small magnetic fields the "perpendicular" strategy proves to be effective even in the long time regime. This can be explained by the fact that, due to memory effects stemming from the interaction with the environment, the coherences experience a revival to the value dependent on $B$ and remain unaffected by phase factors of a type $\exp[i g\mu_{B}Bt]$, which for a non-zero variance $\Delta^{2}B_{prior}$ would lead to their decay. As it was previously observed \cite{Mazurek2014}, the revival of coherences remains in a connection with a revival of rescaled geometric discord of a singlet state of a 2 qubit quantum dot system.   

For prior distributions with $B_0=0$, the evolution of the occupation levels can not be exploited in order to gain information about the magnetic field, due to the symmetry $A(B,t,\alpha)=A(-B,t,\alpha)$ stemming from maximal mixedness of the initial state of environment. Therefore, while in the regime of very small fields we still observe two minima, the origin of the second one can be explained only by memory effects in which coherences are injected back to the system from the environment.  

Clearly, apart from the mentioned minor memory effects, the long time regime is entirely classical, as the estimation there is purely statistical, while in the short time regime, quantum coherences are crucial. For this reason, as we will see further for more particles in non-negligible magnetic field, only for low times entanglement will lead to an enhancement of the estimation.\\

\begin{figure}[h]
\includegraphics[scale=0.74]{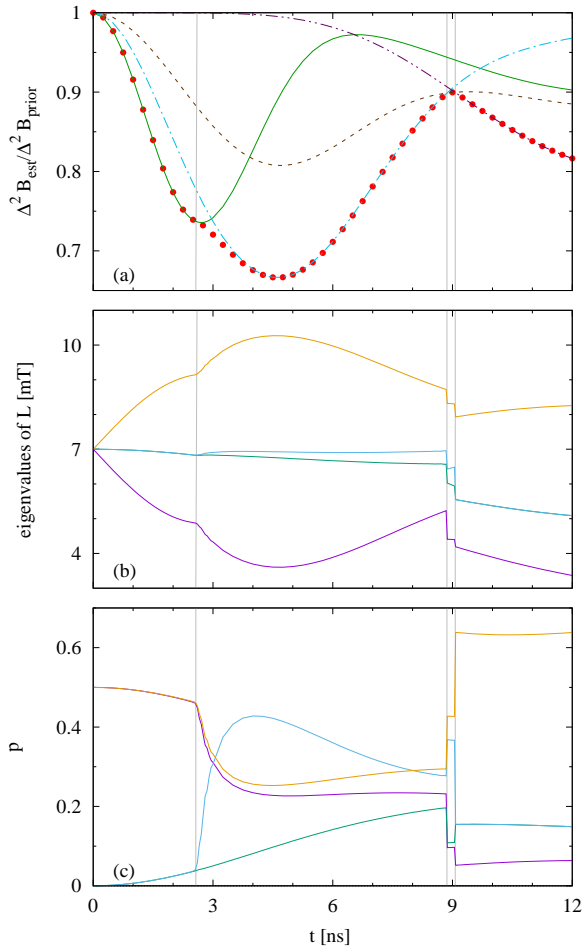}
\caption{ \label{fig:obserwable} 
System of $N=2$ dots, with prior distribution given by $B_{0}$=7 mT, $\Delta B$=4 mT.
(a) Relative precision gain for various initial states: GHZ(2) (green solid line), $a$GHZ(2)$+b|+\>|+\>$, with $a=0.07071$ and  $b=0.63640$ (blue dot-dashed line), $|0+\>$ (grey dashed line), $|++\>$ (purple double-dot-dashed line). Points correspond to the optimal strategy; (b) Eigenvalues of the optimal observable $L$; (c) Probabilities of their measurement.} 
\end{figure}

{\it Two particles (N=2).}  Before we go to more complex systems, it is instructive to focus on two quantum dots, as this will give a hint 
on the general behavior. We consider a Gaussian prior with $B_0=7$ mT and $\Delta B=4$ mT, as for these values we have very well separated two minima and can analyse the transition between the two regimes. Actually, in the case of two particles one can single out several regimes, still clearly separated from each other. The regimes are presented in Fig. \ref{fig:obserwable}. 
The Fig. \ref{fig:obserwable}a
represents corrections given by the optimal strategy as well as optimal corrections to some guessed initial states, while Fig. \ref{fig:obserwable}b
and Fig. \ref{fig:obserwable}c show eigenvalues of the optimal observable, and probabilities of obtaining the corresponding outcomes
for the optimal state, respectively. Sharp transitions between four regions are clearly visible in these latter figures. 
In the first region the optimal state is the GHZ(2), defined as GHZ(N)$=\frac{1}{\sqrt{2}}\big(|0\>^{\otimes N}+|1\>^{\otimes N}\big)$. At about 2.2 ns, phase damping becomes too strong for GHZ(2) state, which causes a transition of the first kind to the second region, where a superposition of $a$GHZ(2)$+b|+\>|+\>$ is optimal, and we use the notation $|+\>=\frac{1}{\sqrt{2}}\big(|0\>+|1\>\big)$. The transition is smooth in a space of the initial optimal states, yet curves describing the evolution of spectrum of $L$ (Fig. \ref{fig:obserwable}b), as well as curves describing probabilities of obtaining measurement outcomes (Fig. \ref{fig:obserwable}c) are non-differentiable. The weight of $|+\>|+\>$ in the optimal superposition increases with selected time of the evolution, reaching approximately the value of $b=0.63640$ (with $a=0.07071$). This marks a transition to the next region, with the optimal initially separable state
$|+\>|0\>$. This region is very short, and soon the final, fourth regime begins, where the state $|0\>|0\>$ becomes optimal. Both previous transitions  are of the zeroth kind -- they are discontinuous in the spectrum of the optimal observable.

In comparison with the single particle case, there are thus two main changes: 
(i) The phase regime has been split into two new regions;
in these two regions, we need initial entanglement, not only superposition. 
(ii) The boundary between the phase and the population regimes has now become a new, intermediate region, where entanglement is not needed, but we still need superposition, although only in one particle state. 

We obtain the following sequence of optimal states:\\
\begin{center}
GHZ $\to a$GHZ$ + b |+\>|+\> \to |+\>|0\> \to |0\>|0\>$.
\end{center}
One should note that in the above $a\neq 0$, which may be partially explained by the fact that GHZ(2) state provides sensibility to $B$ for longer times due to ability to sense it through the measurement of the occupation levels (green line in Fig. \ref{fig:obserwable}a). This is due to GHZ high initial imbalance on the diagonal.   

The situation can be summarized as follows. We can distinguish three main regimes: (1) Regime of initially entangled states, with (1a) Regime of GHZ(2) and (1b) Regime of GHZ(2) superposed with $|+\>|+\>$; (2) Intermediate regime of product coherent states (optimal $|+\>|0\>$); (3) Regime of product states without coherences.

In order to fully describe the optimal measurement strategy for different time regimes, below we move to a description of the optimal observables.
In the (1a) region, the measurement strategy is based on projections onto $e^{\pm \frac{i \pi}{4} \big(S^{z}\otimes \mathbb{1}+\mathbb{1}\otimes S^{z}\big)}$ GHZ(2) states that evolve according to a Hamiltonian $g\mu_{B} B_{0}S^{z}$. This basic metrologic scenario of a parameter sensing via a parameter-dependent phase is efficient for low times, when the evolution is close to unitary, and probabilities of projecting the system at time $t$ into $e^{-2ig\mu_{B} B_{0}S^{z}t} e^{\pm \frac{i \pi}{4} \big(S^{z}\otimes \mathbb{1}+\mathbb{1}\otimes S^{z}\big)}$ GHZ(2) are close to $1/2$ (see Fig. 3c). However, for longer times in the regime, these probabilities decay in favour of likelihood of obtaining two other projections: $|01\>\<01|$ and $|10\>\<10|$. Due to maximal mixedness of environment and symmetry in the initial state of the system, the probabilities remain partially degenerate. 

In the regime (1b), the GHZ(2) ceases to be optimal both because of the external noise and initial lack of knowledge described by a non-zero $\Delta^{2}B_{prior}$, which results in an inability to extract information about the magnetic field from pure phase measurements for longer times. The significance of the second mechanism in limiting magneto-detection is illustrated by the fact that a phase-based magnetometry for a unitary evolution and the initial GHZ(N) state leads to a constant optimal reduction factor $\frac{\Delta^2 B_{est}}{\Delta^2 B_{prior}}$, achieved at times $t_{opt}\propto \frac{1}{N\Delta B}$ (which, for parameters of Fig. \ref{fig:obserwable}, is $t_{opt}\approx$ 3.3 ns); for longer times the reduction factor increases abruptly to $1$ (no information about $B$ is provided). The optimal measurement scheme consists of projections that depend both on diagonal and outer-diagonal elements of a density matrix. It is in this, phase-population strategy mixing regime, that the global minimum of $\frac{\Delta^2 B_{est}}{\Delta^2 B_{prior}}$ is achieved thanks to a prolonged field-dependency of $\Lambda_{B,t}(|++\>)$ through the phase, as well as a field-dependency of $\Lambda_{B,t}($GHZ(2)$)$ both through the phase and occupation levels. In the regime (2) the structure of measurements becomes more transparent, with two projections into the occupation levels and two relying purely on phase-sensing, while in the regime (3), the optimal strategy is to extract information about magnetic field solely from measurements of occupation levels. This, altogether with the aforementioned symmetry of the environment and the structure of the initial state, leads to a degeneration in the spectrum of $L$. \\

\begin{figure}[h]
\includegraphics[scale=0.74]{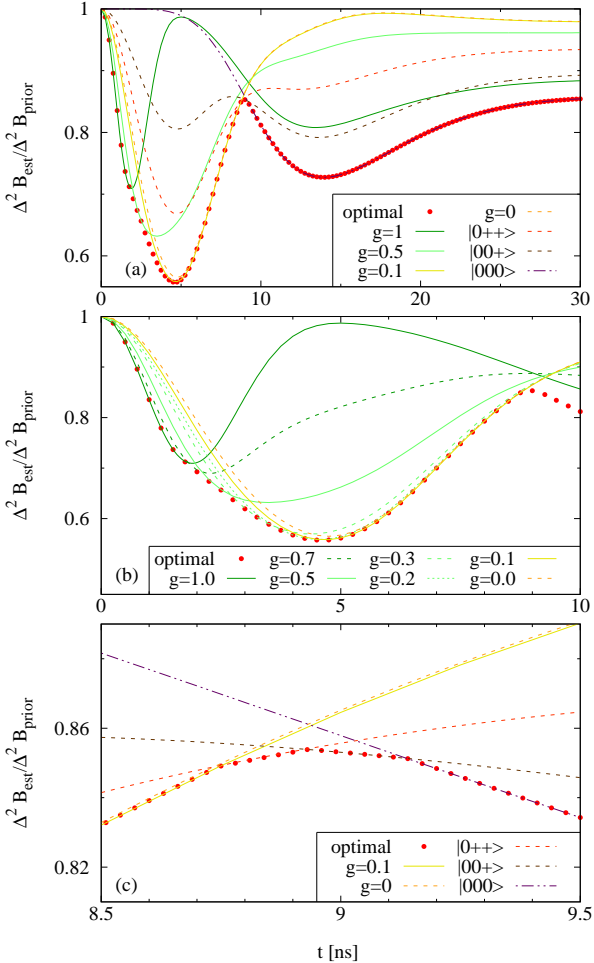}
\caption{ \label{fig:optimal states} 
Relative precision gain for various initial states, for N=3 dots, $B_{0}$=7 mT, $\Delta B$=4 mT. Red dots correspond to the optimal strategy. (a) Optimal measurement strategies for the initial product states and for states of the form  $n$(g GHZ(3)+(1-g)$|$+++$\>$), with normalizing factor $n$. (b) Transition from strategies based on GHZ(3) to those based on its superposition with $|$+++$\>$ states. (c) Transition from strategies based on product coherent states to strategies based on product states without coherences.} 

\end{figure}

\begin{figure}[h]
\includegraphics[scale=0.63]{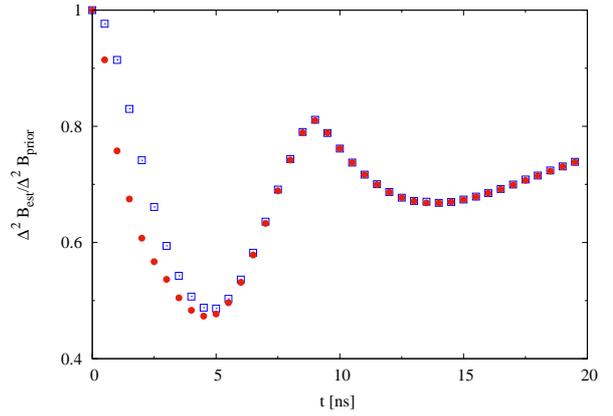}
\caption{\label{fig:biseparable} Relative precision gain for optimal states (red dots) and optimal product states (blue squares) for N=4 quantum dots, $B_{0}$=7 mT, $\Delta B$=4 mT.}\end{figure}

{\it Many particles: enhancement from entanglement}. 
In the next step, we move to systems of $N>2$ quantum dots. We find here the enriched dynamics of the optimal preparation and measurement strategies, that is naturally build on previously explained mechanisms of magneto-detection. A comparison between Fig. \ref{fig:optimal states}a and Fig. \ref{fig:obserwable}a shows that for $N=3$ it is possible to achieve a better correction in the global minimum associated with the initial state in a superposition of GHZ(N) and $|+\>^{\otimes N}$. Fig. \ref{fig:optimal states}b depicts the smoothness in the transition between optimal states for different times in this regime and underlines the necessity of the presence of entanglement -- granted by non-zero presence of GHZ-term in superposition. Monte-Carlo simulation, by which the correction was calculated for randomly selected product initial states, corroborates that entanglement is the necessary resource for achieving this minimum. This is visible for $N=4$ quantum dot systems in Fig. \ref{fig:biseparable}. 

A growing structural complexity of the region that relies on product coherence states is a characteristic feature for a transition into larger systems. Here, all possible combinations of qubits prepared in $|0\>$ and $|+\>$ states (except from $|+++\>$) prove to be optimal, with higher times favoring less coherence present in the initial state.
Fig. \ref{fig:optimal states}c shows the time-dependence of corrections for the series of transitions of zeroth kind. 

\begin{figure}[h]
\includegraphics[scale=0.63]{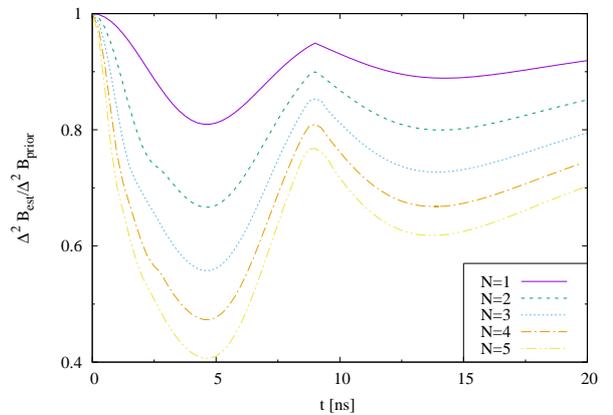}
\caption{ \label{fig:porownanieN} 
Comparison of relative precision gains for different number of dots $N$, for prior distribution given by $B_{0}=7$ mT, $\Delta B=4$ mT.} 
\end{figure}

One should note that precision of the estimation of the field grows with increasing $N$ for all possible times at which measurements are performed (see Fig. \ref{fig:porownanieN}). The justification is the following: for time regimes with optimal GHZ(N) state, the evolution (being close to a unitary one) is affected by increasing $N$ mainly through a change of time scale, as $t_{opt}\propto \frac{1}{N}$. On the other hand, the coherent state $|+\>^{\otimes N}$ structure, largely responsible for magneto-sensitivity of the setup in the neighborhood of the global optimum, for a unitary evolution of systems with increasing $N$ gives the improving optimal correction at times non-dependent on $N$. Therefore, we can postulate that for moderate values of $N$, the GHZ(N)-based range of applicability will decrease in favor of a superposition regime. The population regime is expected to give increasing sensitivity to the field purely due to an increasing number of values one can attribute to an increasing number of measurements, by which the field can be probed more precisely. In contrast to the entanglement regime, one does not need to take into account effects connected with lack of initial knowledge described by non-zero $\Delta^{2}B_{prior}$; simultaneously, the physical noise plays only positive role. Note that the whole regime is absent for a unitary evolution, which implies lack of transitions of zeroth kind. The strategy based on initially product state with coherences is going to enjoy effects of both mechanisms, achieving increasing sensitivity with growing $N$. One should note that product state strategy relies entirely on local measurements.       

\begin{figure}
\includegraphics[scale=0.63]{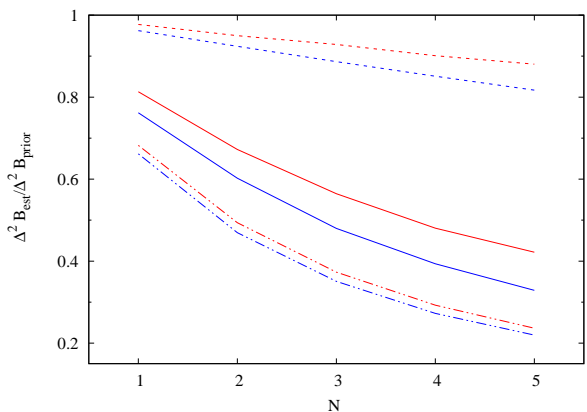}
\caption{ \label{fig:porownanieoptimum} 
Comparison of relative precision gains at optimal times, for different number of dots $N$, dependent on priors: $\Delta B =1$ mT (dashed lines), $\Delta B =4$ mT (solid lines), $\Delta B =10$ mT (dot-dashed lines), $B_{0}=0$ mT (red lines), $B_{0}=1$ T (blue lines). For a given $\Delta B$ and $0<B_{0}<1$ T, the corresponding line lies between the corresponding red and blue lines. Discrete data points are connected by lines for visual transparency.} 
\end{figure}

To fully demonstrate the effects of increasing $N$ on field-probing sensitivity, in Fig. \ref{fig:porownanieoptimum} we illustrate corrections at the global minimum for prior Gaussian distributions with different parameters. 

\section{DISCUSSION}\label{sec:discussion}

We have performed a Bayesian magnetometry analysis for a Gaussian prior distribution of the magnetic field for small (N=1-5) systems of electron spin quantum dots independently interacting with nuclear spin environments in maximally mixed states. The quantitative picture showing sharp transitions of both the optimal initial states (except the first transition point), as well as of the optimal measurements has been identified. 

The standard situation considered in the literature is when the parameter under consideration (here – the magnetic field) is encoded into the system {\it directly} and the noise can only destroy that information.  Here the dynamics makes the parameter imprinted both on the system and environment or –- strictly speaking -– into a global state of both.  Despite the fact that the initial ancillas are maximally noisy and that the final noisy dynamics acts here completely locally, the corresponding noise is unavoidably ,,convoluted'' with the original dynamics and the final result is such that we get the product noisy dynamics which has the parameter imprinted in a {\it nonstandard}, nonlinear way. On the other hand the imprinting of the magnetic field by unitary dynamics is restricted to the Bloch sphere. Effectively we have then the two scenarios. 

In the latter the parameter is imprinted in the states on the sphere, while in the former, it is imprinted in the mixed states that in general belong to the interior of the sphere. It seems that this is {\it the geometry} of the two sets out of which only the one has the nonzero volume, that in general might make the difference in favor of the noisy scenario. 

There are other questions that one can ask here.
First, to what extent the cascade structure of optimal states, from GHZ state to a product of eigenstates of Pauli $\sigma_{z}$ operator, is a universal phenomenon?
Second, is it possible to find a quantum model where
the effect of memory is more visible?
The important problem is also the optimal increase
of information gain in Bayesian approach, which corresponds
to the question how the minimum of the variance ratio behaves in the limit of large N.
This is the problem of high numerical complexity.
Finally, let us note that in our model, local Hamiltonians of
the environments are absent. This is because Zeeman splittings of
the nuclear spins are much smaller than that of the electron spin. It makes the effects stemming from the selfhamiltonians negligible for time scales under the consideration. On the other hand, if we had another physical system with comparable time scales for all subsystems  - the ones that we can initiate and
measure and the ones that are not controlled (environment),
then it may happen that the presence of the local hamiltonians of the
environment leads to better imprinting of the field into the final state
of the measured system. This would require a separate research on
another physical model.

\section*{ACKNOWLEDGMENTS} 
We would like to thank Rafa\l{} Demkowicz-Dobrza\'nski for discussions and  Piotr Gnaci\'nski for helpful assistance regarding numerical simulations. This work was supported by Polish National Science
Centre project 2011/01/B/ST2/05459. \L{}. Cz. acknowledges the support from ERC AdG QOLAPS. Numerical calculations were performed in Academic Computer Centre in Gdansk (CI TASK).

\end{document}